\documentclass[aps,prl,twocolumn,groupedaddress]{revtex4}

\begin{document}

\title{Perturbative Formulation and Non-adiabatic Corrections in
Adiabatic Quantum Computing Schemes}

\author{Yu Shi}

\affiliation{
Cavendish Laboratory,
University of Cambridge, Cambridge CB3 0HE, United Kingdom}

\affiliation{Department of Physics, University of Illinois at
Urbana-Champaign, Urbana, IL 61891}

\affiliation{Institute for Theoretical Physics, University of
Cologne, 50937 Cologne, Germany}

\author{Yong-Shi Wu}

\affiliation{ Department of Physics, University of Utah, Salt Lake
City, UT 84112}

\begin{abstract}

Adiabatic limit is the presumption  of the adiabatic geometric
quantum computation and of the  adiabatic quantum algorithm. But
in reality, the variation speed of the Hamiltonian is finite. Here
we develop a general  formulation of adiabatic quantum computing,
which accurately describes the  evolution of the quantum state in
a perturbative way, in which the adiabatic limit is the
zeroth-order approximation. As an application of this formulation,
non-adiabatic correction or error is estimated for several
physical implementations of the adiabatic geometric gates. A
quantum computing process consisting of many adiabatic gate
operations is considered, for which the total non-adiabatic error
is found to be about the sum of those of all the gates. This is a
useful constraint on the computational power. The formalism is
also briefly applied to the adiabatic quantum algorithm.
\end{abstract}

\pacs{03.67.Lx, 03.65.Vf, 74.81.Fa}

\maketitle

Recently  a considerable amount of attention has been paid to the
idea of using geometric phases accumulated by an adiabatically
time-dependent Hamiltonian to realize quantum
gates~\cite{lit1,lit2,lit3}. Construction of universal gates by
geometric quantum teleportation was studied, with the analysis of
errors from imperfect control~\cite{ellinas}. On the other hand,
adiabatic evolution is also the basis of the so-called adiabatic
quantum algorithms~\cite{farhi}, for which the speed and the
overall time have been analyzed~\cite{vandam}.

For these quantum computing schemes to work, it was supposed that
the adiabatic limit is retained. However, in practice, and
particularly in the case of quantum computation, where the
advantage lies in speedup and the operation time should be shorter
than the decoherence time, the evolution is required to be
completed in a finite period of time.  Therefore, it is important
to know the full picture of the evolution of the quantum state and
the non-adiabatic correction, which gives rise to error if the
adiabatic limit is necessary  for the designed quantum computing
scheme. Here we develop a general formulation of adiabatic quantum
computing, applicable to the previously proposed quantum computing
schemes and to any slowly varying Hamiltonian. As an adiabatic
perturbation theory, it accurately describes the quantum evolution
in a perturbative way, in which the adiabatic limit is the
zeroth-order approximation.  As an application, an examination is
made on the non-adiabatic errors in several previously proposed
implementations of the adiabatic geometric gates. We also
investigate the non-adiabatic error in an entire quantum computing
process consisting of many adiabatic gates, which has not been
considered previously. Finally we briefly discuss the adiabatic
quantum algorithm, noting that such an algorithm can still be
implemented even if the non-adiabatic correction is not
vanishingly small.

If the evolution of a time-dependent Hamiltonian is  sufficiently
slow, the adiabatic theorem tells  that {\em in the adiabatic
limit} and under such conditions as continuity, non-crossing and
differentiability,  an instantaneous eigenstate at an initial time
evolves  to a state close to the corresponding instantaneous
eigenstate at a later time~\cite{book}.

In general, using the instantaneous eigenstate
$|\phi_n(t)\rangle$, one can always expand the state of the system
$|\psi(t)\rangle$ as
\begin{equation}
|\psi(t)\rangle = \sum_n a_n(t)|\phi_n (t)\rangle
\exp[i\eta_n(t)],  \label{t}
\end{equation}
where $\eta_n(t)= -\frac{i}{\hbar} \int_{0}^{t} E_n(\tau)d\tau$ is
the dynamic phase. Then  the Schr\"{o}dinger equation
$i\hbar\partial_t |\psi(t)\rangle = H(t)|\psi\rangle$
 leads to
\begin{equation}
\partial_t a_n(t) =
- \sum_m a_m(t) \langle \phi_n(t)|\partial_t \phi_m(t)\rangle
\exp [i\eta_m(t)-i\eta_n(t)], \label{dif}
\end{equation}
which, together with the initial condition $a_n(0) \equiv \langle
\phi_n(0)|\psi(0)\rangle$, determines $|\psi(t)\rangle$.

First suppose  $|\psi(0)\rangle$ is a non-degenerate eigenstate
$|\phi_{n}(0)\rangle$. Then {\em in the adiabatic limit},
 one obtains $|\psi(t)\rangle
\approx |\phi_{n}(t)\rangle \exp[i\gamma_n( t)+i\eta_n(t)]$, where
$\gamma_n(t)= \int_{0}^{t}\langle
\phi_{n}|\partial_{\tau}\phi_{n}\rangle d\tau = \int\langle
\phi_{n}|\partial_{\mu}\phi_{n}\rangle dx^{\mu}$ is the geometric
or Berry phase~\cite{berry}.

However, when non-adiabatic correction is considered, the exact
state should be the solution of Eq.~(\ref{dif}). For a slowly
varying $H(t)$, using a perturbative approach, one can obtain
\begin{equation}
\begin{array}{c}
U(t)|\phi_{n}(0)\rangle = \exp [i\gamma_n(t)+i\eta_n(t)]
[|\phi_{n}(t)\rangle + \\
\hbar\sum_{m\neq n}
\frac{|\phi_m(t)\rangle\langle\phi_m(t)|\partial_t\phi_{n}(t)\rangle}
{E_m(t)-E_{n}(t)}]+\cdots. \end{array}\label{1st}
\end{equation}

In general, as in quantum computing,  $|\psi(0)\rangle = \sum_n
a_n(0) |\phi_{n}(0)\rangle$ is a superposition of different
eigenstates. Then linearity of quantum evolution implies
\begin{equation}
|\psi(t)\rangle = \sum_n a_n(0) U(t)|\phi_{n}(0)\rangle,
\label{gen}
\end{equation}
where each $U(t)|\phi_{n}(0)\rangle$ is as given in
Eq.~(\ref{1st}).  Therefore, $a_{n}(t) exp[i\eta_n(t)] =
\sum_{m}\langle \phi_n(t)|U(t)|\phi_{m}(0)\rangle  a_m(0)$. From
(\ref{1st}), one obtains
 $\langle \phi_n(t)|U(t)|\phi_{n}(0)\rangle \approx \langle
\phi_n(t)|U^{(0)}(t)|\phi_{n}(0)\rangle = \exp
[i\gamma_n(t)+i\eta_n(t)]$, while for $n \neq m$,
$\langle\phi_n(t)|U(t)|\phi_{m}(0)\rangle \approx \langle
\phi_n(t)|U^{(1)}(t)|\phi_{m}(0)\rangle  = \hbar\exp
[i\gamma_m(t)+i\eta_m(t)]
\langle\phi_n|\partial_t\phi_{m}\rangle/[E_n(t)-E_{m}(t)]$. Here
$U^{(k)}$ refer $k$th order term. Since $E_n(t) \neq E_m(t)$,
$\eta_n(t) \neq \eta_m(t)$. If one implements an all-geometric
gate, in which the instantaneous basis states are
$|\phi_{n}(t)\rangle$ and $|\phi_{m}(t)\rangle$, the difference
between $\eta_n(t)$ and $\eta_m(t)$ needs to cancelled out by
using a certain method~\cite{lit1}.

In the presence of degeneracy of eigenstates, denote the
eigenstates as $|\phi^n_{\alpha_n}(0)\rangle$, where $n$ labels
the energy levels, while $\alpha_n$ labels the different
eigenstates in the subspace $n$. As the generalization of
Eq.~(\ref{1st}), we obtain
\begin{equation}
\begin{array}{c}
U(t)|\phi^n_{\alpha_n}(0)\rangle = \exp [i\eta_n(t)]
[|\chi^{n}_{\alpha_n}(t)\rangle + \\
\hbar\sum_{m\neq
n}\sum_{\beta_m}
\frac{|\chi^m_{\beta_m}(t)\rangle\langle\chi^m_{\beta_m}(t)|
\partial_t\chi^{n}_{\alpha_n}(t)\rangle}
{E_m(t)-E_{n}(t)}]+\cdots, \end{array}\label{deg}
\end{equation}
with $|\chi^n_{\alpha_n}(t)\rangle = \sum_{\beta_n}
V^n_{\alpha_n\beta_n}(t)|\phi^n_{\beta_n}(t)\rangle$,
$V^n(t)=P\exp\int_{0}^{t} A^n(\tau) d\tau$, where
$A^n_{\alpha_n\beta_n} \equiv \langle \chi^n_{\beta_n}(\tau)|
\partial_t\chi^n_{\alpha_n}(\tau)\rangle$ is the
connection in the  subspace $n$. $V^n$ may be called non-abelian
geometric phase or Wilczek-Zee (WZ) phase~\cite{wz}. In the zeroth
order, $U(t)$ is block-diagonal, each block being a WZ phase in
the subspace of a set of degenerate eigenstates. {\em In the
adiabatic limit}, as a unitary transformation, a non-abelian
geometric phase, i.e. the first term in (\ref{deg}), may be used
to realize a quantum gate~\cite{lit2}.

With the existence of degeneracy of eigenstates, a general
superposition state can be written as
\begin{equation}
|\psi(t)\rangle = \sum_n \sum_{\alpha_n}
a^n_{\alpha_n}(t)|\chi^n_{\alpha_n} (t)\rangle \exp [i\eta_n(t)]. \label{tt1}
\end{equation}
By choosing an appropriate basis for each degenerate subspace, the
initial state can always be expanded in such a way that its
projection in each degenerate subspace is a single eigenstate
$|\phi^n_{\beta_n} (0)\rangle$, i.e.
$|\psi(0)\rangle = \sum_n a^n_{\beta_n}(0)|\chi^n_{\beta_n}
(0)\rangle$,
with $|\chi^n_{\beta_n} (0)\rangle = |\phi^n_{\beta_n}
(0)\rangle$. Therefore
\begin{equation}
|\psi(t)\rangle = \sum_n
 a^n_{\beta_n}(0)U(t)|\chi^n_{\beta_n} (0)\rangle. \label{ttt}
\end{equation}
Thus $a^{n}_{\alpha_n}(t) exp[i\eta_n(t)] = \sum_{m}\sum_{\beta_m}
\langle\chi^n_{\alpha_n}(t)|U(t) |\chi^m_{\beta_m} (0)\rangle
a^m_{\beta_m}(0)$. From (\ref{deg}),
$\langle\chi^n_{\alpha_n}(t)|U(t) |\chi^n_{\beta_n} (0)\rangle
\approx \langle\chi^n_{\alpha_n}(t)|U^{(0)}(t) |\chi^n_{\beta_n}
(0)\rangle = \exp[i\eta_n(t)] \delta_{\alpha_n\beta_n}$, while for
$n \neq m$, $\langle\chi^n_{\alpha_n}(t)|U(t) |\chi^m_{\beta_m}
(0)\rangle\approx  \langle\chi^n_{\alpha_n}(t)|U^{(1)}(t)
|\chi^m_{\beta_m} (0)\rangle  = \hbar \exp[i\eta_m(t)]
\langle\chi^n_{\alpha_n}(t)|\partial_t\chi^{m}_{\beta_m}(t)\rangle/[
E_n(t)-E_{m}(t)]$.

Through this formulation, it becomes clear that the adiabatic
quantum computing is based on $\langle
\phi_n(t)|U^{(0)}(t)|\phi_{n}(0)\rangle$ or
$\langle\chi^n_{\alpha_n}(t)|U^{(0)}(t) |\chi^n_{\beta_n}
(0)\rangle$, with higher-order terms neglected. Besides,  while
the previous proposals of adiabatic geometric gates are based on
closed paths, there is nothing in principle against using open
paths, as far as the corresponding geometric phases can be
detected~\cite{wu}. Another noteworthy point, which was not
pointed out before, is that when the qubits under a gate operation
is entangled with other qubits, the linearity of quantum evolution
guarantees that the gate operation is still given by
Eq.~(\ref{1st}) or (\ref{deg}), where the eigenstates are those of
this concerned gate; one may include in the coefficients $a_n(0)$
or $a_{\alpha_n}^n(0)$ the states of the other qubits projected in
the same branch as the eigenstates of the gated qubits. This is
crucial for the possibility that different adiabatic geometric
gates can be networked.

There is a significant difference in the uses of (abelian) Berry
phase and WZ phase to realize a quantum gate, under adiabatic
limit. For a Berry phase gate, it is necessary to have $d$
non-degenerate states, where $d$ is the Hilbert space dimension of
the gate. For a WZ phase gate, one intentionally restrict the gate
in a single degenerate eigenspace. A quantum gate based on WZ
phase is more advantageous than that based on Berry phase, on the
aspect that for the former, in the adiabatic limit, the state is
always an instantaneous eigenstate of the Hamiltonian, hence there
is no dynamical phase difference between the basis states,  and it
is more stable against environmental perturbation.

The  non-adiabatic correction or error at time $t$
 is $\epsilon (t) \equiv [U(t)-U^{(0)}(t)]|\psi(0)\rangle
=\sum_{k=1}^{\infty} U^{(k)}(t)|\psi(0)\rangle \approx
U^{(1)}(t)|\psi(0)\rangle$.  The adiabatic limit means
$|\epsilon(T)| \ll 1$. For  a Berry phase  gate, with
$|\psi(0)\rangle = \sum_n a_n(0) |\phi_n(0)\rangle$,
\begin{equation}
\epsilon (t) \approx   \sum_{n} a_n(0) \sum_{m \neq n} \langle
\phi_m(t)|U^{(1)}(t)|\phi_{n}(0)\rangle
 |\phi_n(0)\rangle. \label{c1}
\end{equation}
For a WZ phase gate  geometric gate  at $E_n$, with
$|\psi(0)\rangle =\sum_{\alpha_n} a_{\alpha_n} (0)
|\phi^n_{\alpha_n}(0)\rangle$,
\begin{equation}
\epsilon (t) \approx   \sum_{\alpha_n} a_{\alpha_n}(0) \sum_{m
\neq n}\sum_{\beta_m}  \langle\chi^m_{\beta_m}(t)|U^{(1)}(t)
|\chi^n_{\alpha_n} (0)\rangle |\phi^n_{\alpha_n}(0)\rangle.
\label{c2}
\end{equation}

Note that  the first order correction at time $t$ is determined
only by eigenvalues, eigenstates and their time derivatives at
$t$, hence is history-independent. This simplifies the analysis.
The time derivatives do depend on the details of time-dependence.
However, since only the path is specified~\cite{cen}, without the
fine control of the dynamics, numerically it suffices to obtain
the order of magnitude. The first-order correction is $ \sim
\hbar/\Delta T$, where $T$ is the time duration of the gate
operation, $\Delta$ is the minimum energy gap with other
eigenstates. It is the presumption of ``slow variation'' or
perturbative approach that $ \hbar/\Delta T < 1$. The $k$-th order
correction is $\sim (\hbar/\Delta T)^k$.

As applications, we now apply the above results to  several
physical implementations previously proposed. The first proposal,
based on Berry phase, uses NMR~\cite{lit1}. The Hamiltonian, in
the rotating frame, is $H(t) = \mathbf{R}(t) \cdot \mathbf{I}$,
where $\mathbf{R}\equiv (R_x, R_y, R_z)=\hbar(\omega_1
\cos\phi,\omega_1 \sin\phi, \omega_0-\omega)$, $
\mathbf{I}=\frac{1}{2}(\sigma_x, \sigma_y,\sigma_z)$, $\omega_0$
is proportional to the static magnetic field in $z$ direction,
$\omega_1$ is proportional to the RF magnetic field in $xy$ plane,
$\omega$ is its angular frequency, $\phi$ is its initial phase.
The instantaneous eigenstates is $|1(t)\rangle =
1/\sqrt{R}[(R_x-iR_y)/\sqrt{R-R_z}|\uparrow\rangle
+\sqrt{R-R_z}|\downarrow\rangle]$,  with eigenvalue $R/2$, and
$|0(t)\rangle =1/\sqrt{R}[(R_x-iR_y)/\sqrt{R+R_z}|\uparrow\rangle
-\sqrt{R+R_z}|\downarrow\rangle]$  with eigenvalue $-R/2$, where
$R \equiv |\mathbf{R}|$. From this, one obtains, for $n=0,1$,
$U_{nn}(0,t)= \exp[i\gamma_n(t)+i\eta_n(t)]$. $\eta_0(t)=Rt/2$,
$\eta_1(t)=-Rt/2$. The Berry phase $\gamma_n(t)$ is,
in the case of a cycle path $C$, $\Omega(C)/2$ for $n=0$ and
$-\Omega(C)$ for $n=1$, where $\Omega(C)$ is the solid angel that
$C$ subtends at $R=0$. It is straightforward to write down
$U_{01}(0,t)$ and $U_{10}(0,t)$. For a gate operation of a period
$T$, the order of magnitude of these two matrix elements, and thus
the non-adiabatic correction, is
$\hbar/RT=1/\sqrt{\omega_1^2+(\omega_0-\omega)^2}T$. The two-bit
gate, of qubits $a$ and $b$,  is effected by addition of the
interaction $\hbar J I_{az}I_{bz}$. For the conditional phases of
qubit $a$,
 $\omega_{a0}$  shifted to
to $\omega_{a0}+ J I_{bz}= \omega_{a0}\pm J/2$, depending on the
basis state $|I_{bz}\rangle$ of $b$.
One can obtain the non-adiabatic corrections in the two subspaces,
with the substitution of $\omega_{a0}\pm J/2$ for $\omega_0$ in
$R$ above. For a gate  as in \cite{lit1}, the gaps are of the
order of several hundred Hertz, while $T$ is of the order of
second, hence the non-adiabatic corrections are  of the order of
$10^{-2}$.

This method was also applied to a Josephson junction
circuit~\cite{falci}. The effective Hamiltonian is still as that
for NMR,  now with $\mathbf{R} = (E_J\cos\alpha, -E_J\sin\alpha,
E_c(1-2n_{off}))$, where $E_J$ and $\alpha$ are decided by the
Josephson couplings of two junctions, $E_c$ is charging energy,
$2e n_{off}$ is the offset charge. In the charging regime, as
used, $E_c \geq E_J$. Thus  the non-adiabatic correction is of the
order of $\hbar/\sqrt{E_J^2+E_c^2(1-2n_{off})^2}T$. Hence if
$1-2n_{off}$ is not too small, the adiabatic condition is
$\hbar/E_cT \ll 1$, more relaxed than previously
thought~\cite{falci}.

An implementation of WZ phase gate was proposed for trapped
ions~\cite{duan,unanyan,recati}. The one-bit gates are based on
the Hamiltonian $H= \hbar |e\rangle(\omega_0\langle
0|+\omega_1\langle 1| + \omega_a\langle a| +h.c.)$,
 One can find that the eigenstates
are: $|\phi^1\rangle =(\omega|e\rangle + \omega_0^*|0\rangle +
\omega_1^*|1\rangle +\omega_a^*|a\rangle)/\sqrt{2}\omega$ with
eigenvalue $\hbar\omega$, where
$\omega=\sqrt{\omega_0^2+\omega_1^2+\omega_a^2}$,
$|\phi^0_{\alpha}\rangle=(\omega_1|0\rangle -
\omega_a|1\rangle)/\sqrt{|\omega_0|^2+|\omega_1|^2}$ and
$|\phi^0_{\beta}\rangle =
(\omega_a\omega_0^*|0\rangle+\omega_a\omega_1^*|1\rangle
-(|\omega_0|^2+|\omega_1|^2)|a\rangle)/
(\omega\sqrt{|\omega_0|^2+|\omega_1|^2})$ with eigenvalue $0$, and
$|\phi^{-1}\rangle =(-\omega|e\rangle + \omega_0^*|0\rangle +
\omega_1^*|1\rangle +\omega_a^*|a\rangle)/\sqrt{2}\omega$ with
eigenvalue $-\hbar\omega$. The WZ phase gates are based on
$U^{00}$, in terms of our notation. Using the instantaneous
eigenstates and eigenvalues,  the non-adiabatic correction is
obtained as $\sum_{n=-1,1}\sum_{x=\alpha,\beta}
a_{x}(0)U^{n0}_{x}(T) |\phi^0_{x}(0)\rangle$, whose order of
magnitude  is of $1/\omega T$.  The two-bit gate proposed there is
only a Berry phase gate  under the Hamiltonian~\cite{duan} $H_{jk}
= \frac{\eta^2}{\delta}[-|\Omega_1|^2\sigma_{j1}^{\phi_1}
\sigma_{k1}^{\phi_1} +|\Omega_a|^2\sigma_{ja}^{\phi_a}
\sigma_{ka}^{\phi_a}]$, where $\sigma_{j\mu}^{\phi_{\mu}} \equiv
e^{i\phi_{\mu}}|e\rangle_{jj}\langle \mu|+h.c.$,
$\phi_1-\phi_a=\phi/2$, using the notations therein. The
eigenstates are $\phi^{1}= (-|\Omega_1|^2e^{-i\phi}|11\rangle
+|\Omega_a|^2|aa\rangle +\sqrt{|\Omega_1|^4+|\Omega_a|^4}
|ee\rangle)/\sqrt{2(|\Omega_1|^4+|\Omega_a|^4)}$
 with eigenvalue $\frac{\eta^2}{\delta}\sqrt{|\Omega_1|^4+|\Omega_a|^4}$,
$|\phi^0\rangle=(|\Omega_a|^2|11\rangle
+|\Omega_1|^2e^{i\phi}|aa\rangle)/
\sqrt{|\Omega_1|^4+|\Omega_a|^4}$ with eigenvalue $0$, and
$\phi^{-1}= (-|\Omega_1|^2e^{-i\phi}|11\rangle
+|\Omega_a|^2|aa\rangle
-\sqrt{|\Omega_1|^4+|\Omega_a|^4}|ee\rangle)/
\sqrt{2(|\Omega_1|^4+|\Omega_a|^4)}$ with eigenvalue
$-\frac{\eta^2}{\delta}\sqrt{|\Omega_1|^4+|\Omega_a|^4}$. It was
proposed to use $|\phi^0\rangle $ to implement the phase gate. The
non-adiabatic correction is of the order of
$\hbar/T(\eta^2/\delta)\sqrt{|\Omega_1|^4+|\Omega_a|^4}$.

Similar proposals were also made in  Josephson junction charge
qubits~\cite{choi,faoro}. For the Hamiltonian  used in
\cite{choi}, there are an  eigenstate with eigenvalue
$\sqrt{h^2+|J_1|^2+|J_2|^2}$, two degenerate eigenstates with
eigenvalue $h$, two degenerate eigenstates with eigenvalue $-h$,
which are used to implement the WZ phase gate, and one ground
state with eigenvalue $\sqrt{h^2+|J_1|^2+|J_2|^2}$, where
$h=E_c(1-2n_{off})/2$.  Thus the non-adiabatic correction is of
the order of $\hbar/(\sqrt{h^2+|J_1|^2+|J_2|^2}-h)T$. In the
two-bit implementation, the eigenvalues are
$-\sqrt{|J_b|^2+(2h)^2}$, $-2h$, $0$, $2h$,
$\sqrt{|J_b|^2+(2h)^2}$. The eigenstates with eigenvalue $-2h$ are
used as the qubit states.  The non-adiabatic correction is of the
order of $\hbar/\Delta T$, where $\Delta$ is the smaller one of
$\sqrt{|J_b|^2+(2h)^2}-2h$ and $2h$. Suppose the order of
magnitude of Josephson energy is $J$. Then if $1-2n_{off}$ is
close to $1$, the energy gap is of the order of $J^2/E_c$ in the
single-bit gate, and is of the order of $J^2/4E_c$ in the two-bit
gate. Since $E_c \geq J$, the energy gap is smaller than $J$.
Hence  compared with the case of \cite{falci},  the adiabatic
condition is harder to meet, i.e. the non-adiabatic correction is
larger. On the other hand, if $1-2n_{off}$ is tuned to be very
small, then the energy gap for the cases of both \cite{choi} and
\cite{falci} are of the order of the Josephson energy.

In the one-bit gate in \cite{faoro}, the energy eigenvalues are
$\delta E_c+\sqrt{(\delta E_c)^2+2J^2}$, $0$, which is with
twofold degeneracy and is used to implement the WZ  gate, and
$\delta E_c-\sqrt{(\delta E_c)^2+J^2}$, where $\delta E_c$ is some
charging energy difference, $J^2=|J_L|^2+|J_M|^2+ |J_R|^2$, using
the notations there. Thus the non-adiabatic correction is of the
order of $\hbar/\Delta T$, where $\Delta$ is the smaller one of
 $|\delta E_c+\sqrt{(\delta E_c)^2+J^2}|$ and
 $|\delta E_c-\sqrt{(\delta E_c)^2+J^2}|$. Hence $\Delta$ is of the order of
 $\delta E_c$ if $\delta E_c > J$, and is of the order of $J$ if
 $\delta E_c < J$.  For the two-bit gate,
the three energy eigenvalues are $\sqrt{|J_X|^2+|J_M^{(2)}|^2}/2$,
$0$, and $-\sqrt{|J_X|^2+|J_M^{(2)}|^2}/2$, where the parameters
are as defined there. Hence the non-adiabatic correction is of the
order of $2\hbar/\sqrt{|J_X|^2+|J_M^{(2)}|^2}T$. For both one-bit
and two-bit gates, the energy gap is at most of the order of
Josephson energy. Therefore, though WZ gate has more advantages
over Berry gate, the non-adiabatic error for \cite{choi,faoro} is
larger than that for \cite{falci}.

A quantum computing process in a gate array consists of many gate
operations on a large number of qubits, hence a complete
estimation of error must include its scaling with the number of
gate operations. Suppose from time $0$ to $T$, $M_1$ adiabatic
gates, denoted as $U^j(t)$, $(j=1,\cdots,M_1)$, are in parallel
operation, each on a small number of (say, one or two) qubits. For
$0 \leq t \leq T$,  the entire quantum computer evolves as
$|\Psi(t)\rangle =U^{1}(t)\cdots U^{M_1}(t)|\Psi(0)\rangle =
 \sum_{i_1\cdots i_{M_1} i_r}
U^1(t)|\phi^1_{i_1}(0)\rangle\cdots
U^{M_1}(t)|\phi_{i_{M_1}}^{M_1}(0)\rangle |\phi^{r}_{i_r}\rangle$,
where $|\phi^j_{i_j}\rangle$ is a basis state of the qubits acted
by the $j$-th gate, $r$ denotes the rest qubits,  which are not
operated by any gate during this period.  We know that $U^j(t)
|\phi^j_{i_j}(0)\rangle = U^j_0(t)|\phi^j_{i_j}(0)\rangle+
\epsilon^j(t)$, where $U^j_0(t)$ represents the  adiabatic limit
of $j$-th gate, while $\epsilon^j(t) <1$ is its non-adiabatic
correction. The state of the quantum computer at $T$ is then
$|\Psi(T)\rangle = |\Psi_0(T)\rangle +\sigma(0,T)$, where
$|\Psi_0(T)\rangle = U^1_0(T) \cdots U^{M_1}_0(T)
|\Psi(0)\rangle$, $\sigma (0,T) \approx \sum_{j=1}^{M_1}
\epsilon^j(T)$ is the first-order error of the entire quantum
computer accumulated from time $0$ to $T$. Afterwards,  during $T
< t <2T$, the quantum computer is operated in parallel by $M_2$
gates, labelled as $M_1+1,\cdots,M_1+M_2$, to which the qubits are
allocated  in a way usually different from the period $0<t< T$.
Using a derivation similar to the above, it can be obtained that
$|\Psi(2T)\rangle = {U}_0^{M_1+1}(T) \cdots {U}^{M_1+M_2}_0(T)
|\Psi(T)\rangle + \sigma(T,2T)$, where $\sigma(T,2T)\approx
\sum_{j=M_1+1}^{M_1+M_2} \epsilon^j(T)$ is the error of the entire
quantum computer accumulated from $T$ to $2T$. Therefore
$|\Psi(2T)\rangle = {U}^{M_1+M_2}_0(T)\cdots {U}_0^{M_1+1}(T)
\cdots U^{M_1}_0(T)\cdots U^{1}_0|\Psi(0)\rangle+\sigma(0,2T)$,
where $\sigma(0,2T)\approx\sigma(0,T)+ \sigma(T,2T) \approx
\sum_{j=1}^{M_1+M_2} \epsilon^j(T)$ is the total error at $2T$.
Therefore, for a quantum computing process consisting of many gate
operations, no matter how they are arranged in space and time, the
total non-adiabatic error is, to the first-order, just the sum of
the errors of all these gates.

Suppose for each adiabatic gate, the time duration  $\leq T$, and
the minimum energy gap with other eigenstates  $\leq \Delta$. Thus
the {\em lower bound} of the non-adiabatic error for each gate is
$||\epsilon(T)|| \sim \hbar/\Delta T$. Hence the lower bound of
the total error is $\sigma \approx M \epsilon(T)$, where $M$ is is
total number of gate operations. For a quantum computing process
to make sense, it is constrained that $||\sigma || < 1$.
Therefore, $M < 1/||\epsilon(T)|| = \Delta T/\hbar$. In Shor's
algorithm, to factor a number $N$, $M \sim 300(\log_{10}
N)^3$~\cite{steane}. Therefore, adiabatic quantum computing can at
most factor $N \approx 10^{(1/300||\epsilon(T)||)^{1/3}}$. For
$||\epsilon(T)||$ of the order $10^{-2}$, $N \approx 10$.

Let us switch to  the adiabatic quantum algorithm~\cite{farhi},
which is based on adiabatically varying the Hamiltonian from a
beginning Hamiltonian $H_b$ at $t=0$ to a final one $H_p$ at
$t=T$.  Under the adiabatic limit, if the system starts with the
ground state of $H_b$, it ends up as the ground state of $H_p$,
which gives the solution to an optimization problem. For a finite
varying rate of the Hamiltonian, to the first order, the more
accurate state is  as given in Eq.~(\ref{1st}). Hence to the first
order approximation, the non-adiabatic correction at time $T$ is
of the order of
 $1/\Delta_pT$, where
$\Delta_p$ is the energy gap of $H_p$, {\em which is independent
of the specific path in which $H_b$ is evolved to $H_p$}.

According to (\ref{1st}),  as far the perturbative approach is
valid, i.e. $|\langle
\phi_0(t)|\partial_t\phi_n(t)\rangle/(E_n-E_0)| < 1$, the
measurement shows that one of the eigenstates appears with
probability clearly the largest. Then one can know that this state
corresponds to the ground state and thus the solution to the
problem. This is consistent with \cite{vandam}.

To summarize, we developed a general, perturbative, formulation of
the adiabatic quantum computing schemes, which perturbatively
describe the accurate evolution of the state. It leads to a deeper
understanding of related issues.  The formalism is applied to
analyze both the the adiabatic geometric quantum computation and
the adiabatic quantum algorithm. The order of magnitude of the
first-order non-adiabatic error is the inverse of the executing
time times the minimum gap with other eigenstates. Several
proposed physical implementations of the former are considered
from this point of view.  Different proposals based on charging
Josephson junctions are compared.  We also consider an entire
quantum computing process consisting of many adiabatic gates,
obtaining the lower bound of the non-adiabatic error, as an
interesting constraint on the power of the quantum computation
based on adiabatic geometric gates. One needs to enlarge the
energy gap in order to reduce the non-adiabatic error and thus
improve the computational power. For the adiabatic quantum
algorithm, it is noted that it can be realized as far as the
perturbative approach, rather than the rigorous adiabatic limit,
is valid, hence the computational time may be appropriately
shortened.

Y.S. was supported by a Humboldt Fellowship when visiting Germany.

\end{document}